# Evolutionary Computation-Assisted Brainwriting for Large-Scale Online Ideation

NOBUO NAMURA

Hitachi, Ltd., Hitachi, Japan, nobuo.namura.gx@hitachi.com

TATSUYA HASEBE

Hitachi, Ltd., Hitachi, Japan

Brainstorming is an effective technique for offline ideation although the number of participants able to join an ideation session and suggest ideas is limited. To increase the diversity and quality of the ideas suggested, many participants with various backgrounds should be able to join the session. We have devised an evolutionary computation-assisted brainwriting method for large-scale online ideation. In this method, participants not only suggest ideas but also evaluate ideas previously suggested by other participants. The evaluation results are used in the evolutionary computation to identify good ideas to which the participants can be exposed via a brainwriting-like interface. We compared the performance of the proposed method with that of a simple online brainwriting method for large-scale online ideation with more than 30 participants. The proposed method enhanced robustness of idea quality improvement due to preferentially exposing the participants to good ideas.

## 1 INTRODUCTION

Brainstorming [1] is an effective technique for generating innovative ideas for future products and services. Brainstorming is typically conducted by gathering approximately 3–10 participants in a meeting room and allowing them to converse and suggest ideas. However, gathering them in one place at the same time is difficult if some are working remotely such as at home or in an overseas office in a different time zone. Various studies on electronic brainstorming have tackled this problem [2–5], and various online ideation platforms have been provided. Studies have shown that electronically exposing participants to ideas previously suggested by the other participants enables them to generate ideas better than they would otherwise [6, 7].

Crowd-sourced ideation has received much attention as it enables a larger number of participants to be gathered and a larger number of ideas to be generated [8–10]. Moreover crowd-sourced ideation enhances the diversity of the participants to include people related to the entire value chain of the products and services. Utilizing the knowledge and inspiration of the participants should help designers and planners create innovative products and services because good ideas can be generated not only by domain experts but also by nonexperts, including customers. In large-scale ideation, however, the participants are unable to scan all of the potentially hundreds to thousands of ideas previously generated and identify ones that might stimulate their thinking. Applying electronic brainstorming to large-scale ideation requires a method for exposing the participants to stimulating ideas. Wang *et al.* demonstrated that exposing people to novel ideas increased the novelty of the ideas they generated compared with without such stimulation and that exposure to common ideas had no effect on novelty [11].

Various methods have been proposed for providing suitable stimulation to participants in large-scale ideation. Chan *et al*. introduced expert facilitation into crowd-sourced brainstorming and showed that it increases quantity and creativity of the ideas suggested [12]. They evaluated the use of automatic sensemaking to extract essential information from numerous ideas and found that it can improve the quantity and diversity of ideas at the cost of mean quality [13]. Tanaka *et al*. compared two types of stimulation methods: exposing participants to two randomly selected ideas versus exposing them to one idea and a criticism of it [14]. Exposing them to two ideas resulted in the generation of ideas with higher mean novelty and practicality whereas exposing them to an idea and its criticism improved the novelty and practicality of the top five ideas. Sakamoto and Bao applied tournament selection, which is usually used in genetic algorithms to select the parents of the next generation, to crowd-sourced ideation [15]. They used three crowds: one crowd generated ideas, another crowd evaluated their quality, and the third crowd combined pairs of computer-selected ideas, which were selected through a tournament method, with a bias toward higher quality.

Brainwriting [16–19] is an effective technique for exposing participants to previously generated ideas to stimulate their thinking. Each participant writes three ideas on a worksheet and passes it to the next participant. That participant reads the ideas on the worksheet to stimulate his or her thinking and then writes three more ideas on the worksheet. Brainwriting is analogous to online ideation as there is no group discussion during the ideation process [20]. Application of online brainwriting to large-scale ideation requires a method for exposing the participants to stimulating ideas.

We have devised an evolutionary computation-assisted brainwriting (ECBW) method for doing this. It combines brainwriting and evolutionary computation represented by the genetic algorithm. The ECBW method is practical for large-scale ideation among people in remote locations and different time zones. The participants are exposed to potentially stimulating ideas selected by interactive evolutionary computation (IEC) within the brainwriting framework. IEC is a typical evolutionary computation method in which an objective function is evaluated by people rather than by computer [21]. The participants not only suggest ideas after being stimulated by exposure to the ideas suggested by the other participants but also evaluate those ideas. The stimulating ideas are selected on the basis of the evaluation scores. Iteration of this process increases idea quality step-by-step. In this study, we compared the performance of the ECBW method with that of a simple online brainwriting (OBW) method by applying them to a large-scale online ideation with more than 30 participants. This comparison showed effects of evolutionary computation on variation in idea quality along with ideation progress. The relationship between ideas suggested by participants and ones to which they were exposed was investigated from quality point of view to reveal process in which good ideas were generated.

## 2 EVOLUTIONARY COMPUTATION-ASSISTED BRAINWRITING

A function for evaluating ideas is provided in the ECBW interface to enable IEC to select ideas on the basis of their evaluation scores. This section describes the interface, the idea database, and the ECBW algorithm as well as the algorithm of the OBW used for comparison.

### 2.1 Interface

As shown in Fig. 1, the ECBW interface consists of three main components in addition to the topic for ideation and the usage instructions.

a. Stimulating idea cells: $3 \times 3$ gray boxes
   b. Vote buttons: $3 \times 3$ check boxes below stimulating idea cells for evaluating ideas
   c. Participant idea cells: red-bordered boxes for participants to enter their own ideas

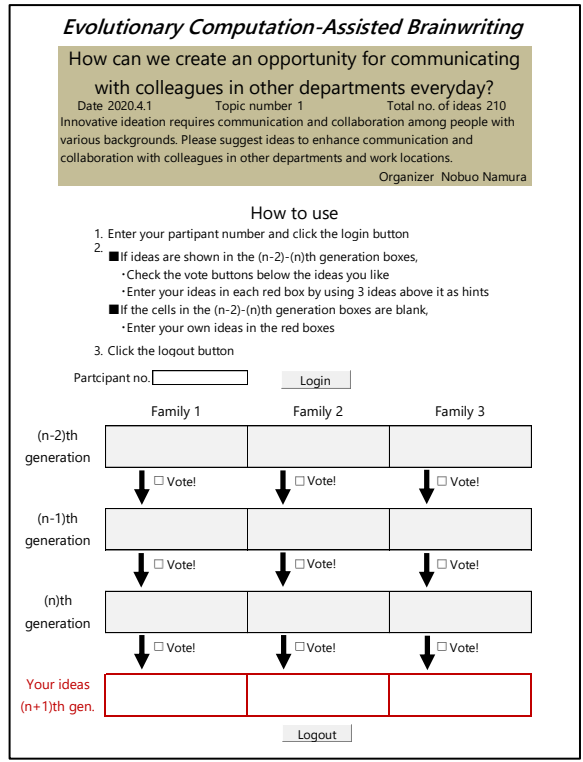

Figure 1: Interface for ECBW: (a) stimulating idea cells, (b) vote buttons, (c) participant idea cells.

Basically, the ECBW algorithm selects $3 \times 3$ ideas to be shown in the stimulating idea cells from the previously suggested ideas on the basis of their scores (the total number of votes received). The participants indicate the ideas they like by using the vote buttons. They then suggest three new ideas on the basis of the three stimulating ideas in each column of cells as well as the brainwriting. Since no ideas have been suggested when ideation begins, the initial participants simply suggest three ideas without any stimulation. Hereafter, these ideas are called "initial ideas."

The ECBW method regards the ideas in each column of cells as a family of ideas. The idea in the participant idea cell is defined as an offspring of the parental idea above it. Ideas belonging to the same family are the posterity of the same initial idea and have the same features inherited from the initial idea. The ECBW method considers $N_f$ families to maintain idea diversity although the interface exposes each participant to only three families selected by the ECBW algorithm.

## 2.2 Database

A database is used to store the ideas to be selected and presented to the participants by the ECBW algorithm, along with their score and family information. Table 1 summarizes the indices of the database and shows example entries. Each idea is stored with its identification (ID), number of presentations, score, participant number, family number, generation number, parent ID, and offspring ID. The ID numbers are assigned in order of idea suggestion. The number of presentations and score represent how many times the idea has been presented and the number of votes it has received. The participant number is the identification number of the participant who suggested the idea. The family number is copied from parent to offspring, so all ideas belonging to the same family have the same family number. Each initial idea is assigned to a unique family, and the family number is the same as the idea ID (*i.e.*, Idea 1 is assigned to Family 1, Idea 2 is assigned to Family 2, and so on). The generation number of an offspring equals that of its parent plus one while the generation numbers of the initial ideas are 1. The parent and offspring IDs are used to compute the total number of presentations and the score for each family. The parent ID of the initial ideas is zero, and ideas may have multiple offspring IDs.

Table 1: Indices and examples of database entries for ECBW.

| ID | ... | 148 | 149 | 150 | ... |
|---|---|---|---|---|---|
| Idea | | "hoge1" | "hoge2" | "hoge3" | |
| No. of presentations | | 5 | 2 | 2 | |
| Score | | 5 | 2 | 1 | |
| Participant no. | | 23 | 23 | 23 | |
| Family no. | | 9 | 3 | 5 | |
| Generation no. | | 7 | 12 | 13 | |
| Parent ID | | 131 | 133 | 111 | |
| Offspring ID | | 151 | 159 | 153 | |

## 2.3 Algorithm

The ECBW algorithm is based on the hypothesis that "even better ideas can be obtained if people are exposed to better stimulating ideas," which is the basis of conventional evolutionary computation. The steps in the ECBW algorithm (Fig. 2) are similar to those in evolutionary computation. First, a participant logs in by entering his or her number into the interface and clicks the login button. This number is used for selecting ideas to be presented. Next, the number of previously suggested ideas $n$ and number of families $N_f$ are compared. If $n < N_f$, initial idea generation is conducted, and the participant is asked to suggest three initial ideas without any stimulation. If $n \geq N_f$, $3 \times 3$ stimulating ideas are selected (three families, with three ideas each) and presented to the participant. First, three families are selected on the basis of their familial score rates (FSRs), which are computed from the total number of idea presentations and the total score for each family. Three ideas are then selected from the ones belonging to each selected family on the basis of their individual score rates (ISRs), which are computed from the number of their presentations and their score. The ideas are then presented to the participant, and the participant is asked to evaluate them by checking the vote buttons below the ideas he/she likes. The number of vote buttons the participant checks is not limited and can range from 0 to 9. The participant then suggests ideas by entering them in the participant idea cells for reproduction (recombination) of ideas and clicks the logout button, which updates the database. Finally, the difference between the updated $n$ and the number of ideas the organizer needs $N$ is compared with the termination condition. The familial and individual selection method and reproduction concept are described next.

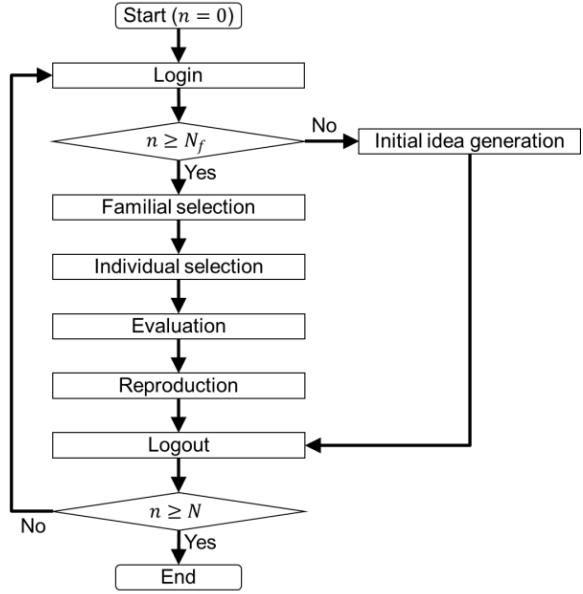

Figure 2: Steps in ECBW algorithm.

### 2.3.1 Familial and individual selection

The families and ideas to be presented are probabilistically selected using the FSRs and ISRs. The FSR of the $i$-th family is defined as

$$FSR_i = \frac{\sum_{j=1}^{J_i} S_{i,j}}{\sum_{j=1}^{J_i} E_{i,j}}, \quad (1)$$

where $E_{i,j}$ and $S_{i,j}$ $(i = 1, 2, \cdots, N_f, \; j = 1, 2, \cdots, J_i)$ are the number of presentations and score for the $j$-th idea in the $i$-th family, respectively. $J_i$ is the number of previously suggested ideas in the $i$-th family. $FSR_i$ is the ratio of the total score to the total number of presentations in the $i$-th family and corresponds to the probability that ideas belonging to the $i$-th family receive votes. In the same manner, the ISR of the $j$-th individual in the $i$-th family is defined as

$$ISR_{i,j} = \frac{S_{i,j}}{E_{i,j}}. \quad (2)$$

$ISR_{i,j}$ corresponds to the probability that participants vote for the idea.

However, $FSR_i$ and $ISR_{i,j}$ are not reliable criteria for selection when $\sum_{j=1}^{J_i} E_{i,j}$ and $E_{i,j}$ are small, which means that $FSR_i$ and $ISR_{i,j}$ depend on the preferences of a small number of participants. In this study, we assumed that $FSR_i$ and $ISR_{i,j}$ computed from less than three presentations were not reliable. Thus, $FSR_i$ and $ISR_{i,j}$ are modified using a correction function, $f(x)$:

$$FSR'_i = \begin{cases} c & if\ E_{i,1} = 0 \\ \frac{f\left(\sum_{j=1}^{J_i} S_{i,j}\right)}{f\left(\sum_{j=1}^{J_i} E_{i,j}\right)} & otherwise \end{cases} \quad (3)$$

$$ISR'_{i,j} = \begin{cases} c & if\ E_{i,j} = 0 \\ 0 & if\ E_{i,j} > 1\ and\ ISR_{i,j} < 0.5 \\ \frac{f(S_{i,j})}{f(E_{i,j})} & otherwise \end{cases} \quad (4)$$

$$f(x) = \begin{cases} \sin\left[\frac{\pi}{2}\left(\frac{x}{a}-1\right)\right] + b & if\ x < a \\ \frac{\pi}{2}\left(\frac{x}{a}-1\right) + b & otherwise \end{cases}, \quad (5)$$

where $FSR'_i$ and $ISR'_{i,j}$ are the modified $FSR_i$ and $ISR_{i,j}$, respectively. The $a$, $b$, and $c$ are parameters used to determine the strength of the correction: $a = 3$, $b = 1.5$, and $c = 2$. $E_{i,1}$ corresponds to the initial idea in the $i$-th family. $FSR'_i$ and $ISR'_{i,j}$ of the families and ideas poorly presented increase from original $FSR_i$ and $ISR_{i,j}$ values. Thus, these families and ideas are preferentially selected and presented to improve the reliability of their $FSR_i$ and $ISR_{i,j}$. By contrast, the second condition in Eq. (4) makes the probability of selecting ideas that are not preferred by participants zero. This condition enables the ECBW method to eliminate nonpreferable ideas at an early stage of ideation and to increase the presentation of preferable ideas and thereby improve the reliability of their $ISR_{i,j}$. Additionally, the $ISR_{i,j}$ of an idea suggested by the current participant is replaced by zero so that the current participant is not exposed to his or her previously suggested ideas. The participant number is used for this.

The probability of selecting the $i$-th family $P_i$ is based on the proportion of its $FSR_i$ to the total $FSR_i$ of all families:

$$P_i = \frac{FSR'_i}{\sum_{i=1}^{N_f} FSR'_i}. \quad (6)$$

In the same manner, the conditional probability that the $j$-th idea is selected given that the $i$-th family is selected is defined as

$$P_{j|i} = \frac{ISR'_{i,j}}{\sum_{j=1}^{J_i} ISR'_{i,j}}. \quad (7)$$

Three different families are iteratively selected with Eq. (6). After one family has been selected, the $FSR_i$ of that family is replaced by zero before the next family is selected and remains zero until the current participant logs out. The same procedure is applied to individual selection. The selected ideas in each family are displayed in the interface in order of their generation; older ideas, which have smaller generation numbers, are shown in the upper stimulating idea cells.

### 2.3.2 Reproduction

In the ECBW algorithm, human creativity is used to reproduce (recombine) ideas. That is, the reproduction is done by the participants. In conventional IEC, the participants only evaluate ideas, and the reproduction is done automatically by a computational algorithm with genetic operators (crossover and mutation). The crossover operator generates new offspring by combining features of multiple ideas while the mutation operator modifies an idea by randomly changing parts of it. The human-based reproduction in the ECBW algorithm is a combination of the crossover and mutation operators because the participants generate ideas on the basis of stimulating ideas (crossover) and adding their own thoughts (mutation). Ideas can be automatically generated by combining words with natural language processing [22] although their quality may not be as good as those that humans generate. Such automatic idea generation requires numerous evaluations by participants to improve the quality of the ideas. This workload can be heavy and unenjoyable for the participants because they simply evaluate ideas without creative process. This is not desirable for a practical ideation system. We use human-based reproduction for enjoyable and efficient ideation since human inspiration is essential for idea generation.

## 2.4 Online brainwriting

The steps in the OBW method are almost the same as those shown in Fig. 2 except for the familial and individual selection. The familial and individual selection in the OBW method does not need the results of the participants' evaluation although the participants are asked to evaluate the ideas through the same interface used for the ECBW method (Fig. 1). The OBW method randomly selects three families, and then the latest three ideas in each selected family are presented to the participant. Hence, almost all previously suggested ideas except for the latest three generations are presented only three times whereas the number of presentations differs from one idea to another in the ECBW method due to its probabilistic selection.

## 3 EXPERIMENTAL SETUP

We compared the $ISR_{i,j}$ and $FSR_i$ of ideas obtained using the ECBW and OBW methods to investigate the effects of evolutionary computation on idea generation. The topic for ideation was the same for both methods: "How can we create an opportunity for communicating with colleagues in other departments everyday?" Each method produced 210 ideas ($N = 210$) including the initial ideas, which were common in both methods, for 12 families ($N_f = 12$). Participants were recruited from the authors' institution and were tasked to use one of the methods on the basis of their institutional ID number: odd or even. Thirty-seven used the ECBW method, and 31 used the OBW method.

The metric for idea quality was $ISR_{i,j}$. Only ideas with $E_{i,j} > 1$ were used for comparison based on $ISR_{i,j}$ due to their reliability although $FSR_i$ was computed from all ideas including ones with $E_{i,j} \leq 1$. In the comparison, the ECBW method stored many ideas with $E_{i,j} = 2$ and $S_{i,j} = 0$ when ideation was completed. This is because these ideas were eliminated from individual selection in the ECBW method and thus had no opportunity for a third presentation. We refer to them as "eliminated ideas."

## 4 RESULTS AND DISCUSSION

### 4.1 Effects of differences in participants and algorithms

The characteristics of ideas produced by the ECBW and OBW methods are shown in Table 2. The difference in the number of participants was negligible because there were no significant differences between the methods (Chi-squared test, $p = .843$) in the proportions of total score to total presentations for all ideas, the "total score rate." The total number of presentations with the ECBW method was lower due to eliminating ideas although the number of ideas produced was the same.

The ECBW method had a significantly lower mean $ISR_{i,j}$ by the Wilcoxon rank sum test ($p < .05$). However, this does not mean that the ECBW method is inferior. It simply reflects the difference in evaluation algorithms. Some ideas were presented more than three times ($E_{i,j} > 3$) with the ECBW method although their $ISR_{i,j}$ could not be higher than 1. When the number of ideas, total score, and total number of presentations were comparable, the mean $ISR_{i,j}$ for the ECBW method, for which some ideas obtained scores larger than 3 ($S_{i,j} > 3$), should be lower than that for the OBW method, for which the scores of all ideas were not larger than 3 ($S_{i,j} \leq 3$).

The idea elimination also reduced the mean $ISR_{i,j}$ in the ECBW method. Figure 3 shows normalized histograms of the $ISR_{i,j}$ for the ECBW and OBW methods. They were generated using four classes as most of the ideas generated with the OBW method had an $ISR_{i,j}$ corresponding to four values: 0/3, 1/3, 2/3, and 3/3. The ideas exactly on the classification boundaries (0.25, 0.5, and 0.75) were counted on both sides of the boundary as 0.5 ideas. Such ideas were frequently obtained in the ECBW method. The histograms show that poor ideas ($ISR_{i,j} \leq 0.25$), including the eliminated ideas, accounted for a large proportion of the ideas generated with the ECBW method. The number of poor ideas was higher with the ECBW method since the eliminated ideas had no opportunity for a third presentation. A third presentation would enable a certain portion of the eliminated ideas to be moved to the next higher class ($0.25 \leq ISR_{i,j} \leq 0.5$). On the other hand, the number of good ideas ($ISR_{i,j} \geq 0.75$) with the ECBW method was also higher than with the OBW although ideas with $E_{i,j} = 2$ and $S_{i,j} = 2$ were included.

Table 2: Characteristics of ideas produced by ECBW and OBW methods.

|  | ECBW |  | OBW |
|---|---|---|---|
| Number of ideas produced | 210 |  | 210 |
| Number of participants | 37 |  | 31 |
| Total no. of presentations | 545 |  | 558 |
| Total score | 221 |  | 223 |
| Total score rate | 0.400 | ≈ | 0.406 |
| Mean $ISR_{i,j}$ | 0.331 | < | 0.397 |

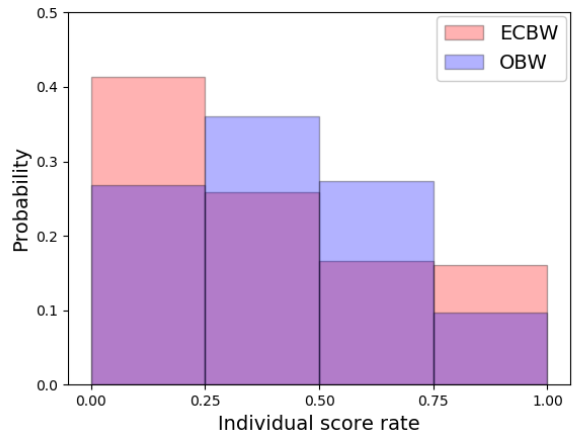

Figure 3: Normalized histograms of individual score rates with ECBW and OBW methods.

### 4.2 Quality improvement in ideas along with ideation progress

According to the hypothesis on which the ECBW is based, the quality of ideas (represented by $ISR_{i,j}$) should increase as ideation progresses. To investigate this, we plotted the mean $ISR_{i,j}$ for every 24 ideas (8 participants) sorted by ID (Fig. 4). The ideas with ID > 168 were ignored because most of them had $E_{i,j} < 2$ with the ECBW method. The horizontal axis in Fig. 4 corresponds to the progression of time. Direct comparison between the methods in the figure is worthless due to the reasons described in subsection 4.1 although the OBW method generally achieved a higher mean $ISR_{i,j}$. Thus, we focus on the variation in the mean $ISR_{i,j}$.

The quality of ideas improved with time until around ID 144 for both methods. Their mean $ISR_{i,j}$ roughly increased up to this point even though no significant difference was found. However, the mean $ISR_{i,j}$ for the OBW method dropped significantly, from 0.5 to 0.3, after this point ($p < .05$). This drop was attributed to successive suggestions of poor ideas, which disrupted the essential concepts inherited from the good ideas. Once three poor ideas are proposed for three successive generations in the same family, the OBW method is unable to present previous good ideas in that family. This is because the OBW method simply presents the three latest ideas. In contrast, the ECBW method can robustly improve the quality of ideas by preferentially presenting good ideas regardless of their generation.

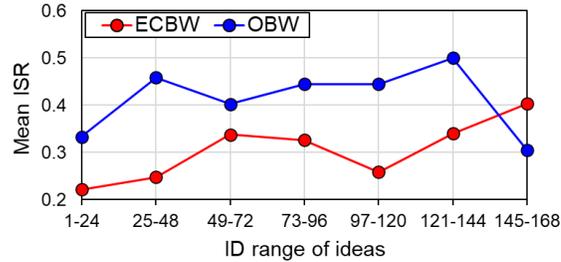

Figure 4: Mean values of individual score rates for every 24 ideas in ECBW an OBW methods.

### 4.3 Relationship between parent and offspring ideas

The relationship between parent and offspring ideas was investigated from the point of view of quality. Figures 5 and 6 show simultaneous probability distributions and transition probability matrices of $ISR_{i,j}$ for the ideas in a parent-offspring relationship, respectively. In both figures, the $ISR_{i,j}$ of the parents and offspring are divided into four classes in the same manner as Fig. 3. Each component in the transition probability matrices indicates the conditional probability of the offspring's $ISR_{i,j}$ given the parent's $ISR_{i,j}$. First, the results with the OBW method are explained to confirm that the basic strategy of ideation follows the hypothesis on which the ECBW method is based. Subsequently, the results with the ECBW method are compared with these of the OBW method to identify its feature.

The OBW method gradually improved idea quality in accordance with the hypothesis. The red-dashed boxes in Figs. 5(b) and 6(b) correspond to the probabilities that the offspring has a one-class higher $ISR_{i,j}$ than the parent. These components had the highest values in each column when the $ISR_{i,j}$ of the offspring was higher than 0.5. The mean $ISR_{i,j}$ of the offspring of a parent with $ISR_{i,j} \geq 0.75$ was significantly higher than that of a parent with $ISR_{i,j} \leq 0.25$ ($p < .05$). Moreover, parents with $ISR_{i,j} \geq 0.5$ more frequently had offspring with $ISR_{i,j} \geq 0.75$ than parents with $ISR_{i,j} \leq 0.5$, as shown by the red-solid boxes in Fig. 6(b). These results support the hypothesis and suggest that good ideas ($ISR_{i,j} \geq 0.75$) are generated by improving the quality in a step-by-step manner, which is an advantage of the OBW method.

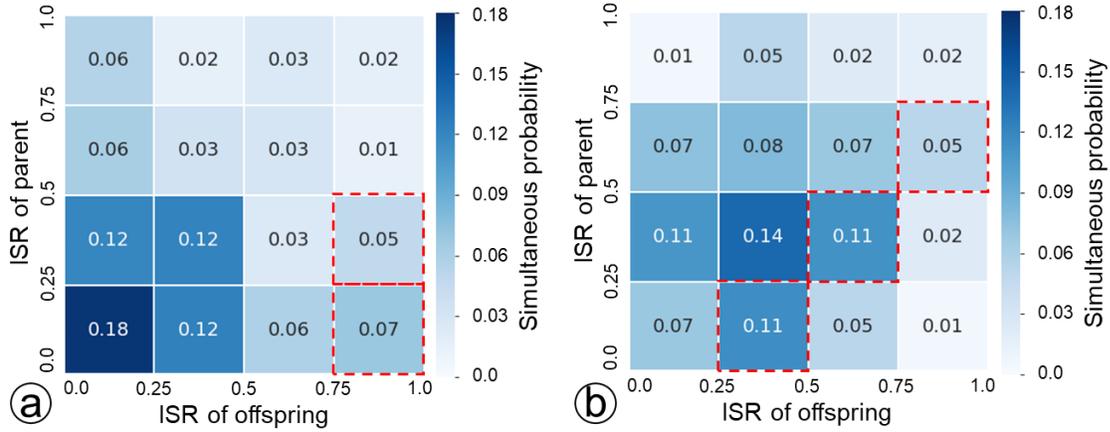

Figure 5: Simultaneous probability distributions of individual score rates for ideas in parent-offspring relationship: (a) ECBW; (b) OBW.

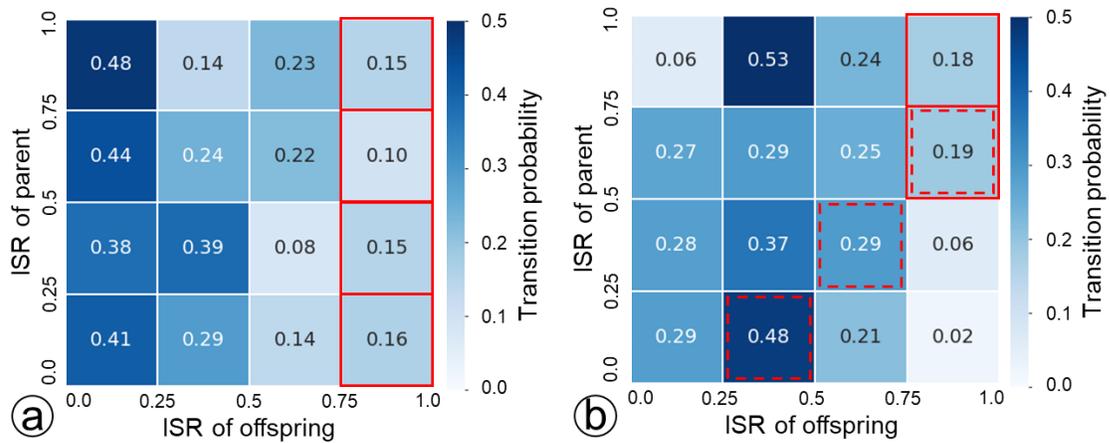

Figure 6: Transition probability matrices of individual score rates for ideas in parent-offspring relationship: (a) ECBW; (b) OBW.

On the other hand, the ECBW method obtained good offspring ($ISR_{i,j} \geq 0.75$) mainly from poor parents, as shown by the red-dashed boxes in Fig. 5(a). This was due to the higher transition probabilities highlighted by the red-solid boxes in Fig. 6(a), which shows that the probabilities of obtaining good offspring were comparable regardless of the parent's $ISR_{i,j}$. The difference between the IE and OBW methods in the transition probability matrices was due to the ideas used to stimulate the participants. The ECBW method tends to show better ideas in the upper cells because the selected ideas are sorted by generation, so later ideas with their smaller number of presentations are shown in the lower cells. The participants were mostly stimulated by the ideas in the upper cells (genetic parents) while the formal parents of the suggested ideas were defined as the ideas in the bottom cells. Thus, good ideas were obtained regardless of the $ISR_{i,j}$ of the formal parent. This advantage enables the ECBW method to robustly improve the quality of ideas, as shown in Fig. 4.

From another point of view, the transition probabilities of an offspring's $ISR_{i,j} \geq 0.75$ given a parent's $ISR_{i,j} \geq 0.5$ with the ECBW method were lower than those with the OBW method. This could be due to the individual selection, in which three stimulating ideas were not always in the parent-offspring relationship, in the ECBW method. These disconnected ideas can make it difficult for participants to identify the common features and essential concepts in each family. One of the participants pointed out this problem after the experiment. For more efficient ideation, the individual selection method should be modified accordingly.

By combining the advantages of the ECBW and OBW methods, a modified individual selection method can be devised. Such a method would utilize the individual selection in the ECBW method for selecting one idea to be shown in the top cell for each family. The other two cells would be filled with the two latest ideas for each family according to the individual selection in the OBW method. Seeing the two latest ideas would enable the participants to gradually improve the quality of ideas while seeing the other idea would stabilize the quality improvement.

### 4.4 Quality of ideas in each family

The differences in idea quality among families were investigated to identify the effects of familial selection and the initial ideas on ideation. Figure 7 shows the $FSR_i$, total number of ideas, number of good ideas ($ISR_{i,j} \geq 0.75$), and proportion of good ideas to total ideas at the end of ideation for each family. The $ISR_{i,j}$ of ideas with $E_{i,j} > 0$ is shown in Fig. 8 by generation number. The closed and open circles correspond to ideas with $E_{i,j} > 1$ and $E_{i,j} > 0$, respectively. Orange circles and dashed lines show branched ideas, which are siblings and cousins of other ideas, in the ECBW method.

As shown in Figs. 7(a, b), there were fewer ideas in the sixth and eighth families than in the other families with the ECBW method because the familial selection reduced their selection probabilities on the basis of $FSR_i$. Reducing the selection probabilities of these families increased the number of presentations of ideas in families with higher $FSR_i$, which enhanced the generation of good ideas.

The initial ideas in the sixth and eighth families were poor or difficult to use for stimulation because these families had a lower $FSR_i$ with both the ECBW and OBW methods. The number of good ideas in these families was 0 and 1, respectively, although the total number of ideas in these families was comparable to those in the other families with the OBW method. Figure 8 also shows that the $ISR_{i,j}$ for these families was especially low until the fifth generation and did not improve much thereafter. Most of the other families had at least one idea with $ISR_{i,j} \geq 0.5$ by the fifth generation due to $ISR_{i,j}$ increasing from the initial ideas.

On the other hand, the initial idea in the fourth family could have been suitable for stimulation since both methods had a large proportion of good ideas in that family. However, the fourth family had fewer ideas than most of the other families whereas it had a comparable $FSR_i$. The familial selection should thus be modified to use another criterion based on the proportion of good ideas instead of $FSR_i$ to enhance the generation of good ideas. The proportion of good ideas itself can be used as a criterion. In this case, correction is needed to define the criterion when there are no good ideas ($ISR_{i,j} \geq 0.75$) at an early stage of ideation.

Comparing the initial ideas in the fourth and eighth families, we see that their $ISR_{i,j}$ was low, as shown in Fig. 8, whereas the quality of the families (represented by $FSR_i$) and the proportion of good ideas (Fig. 7) were substantially different. These results mean that the $ISR_{i,j}$ of ideas and their suitability for stimulating ideas are not always correlated. If the suitability for stimulation can be quantified and used for individual selection, the proportion of good ideas should increase. Using the $ISR_{i,j}$ of an offspring as an individual selection criterion for its parent is one approach to suitability quantification. Definition of the parent-offspring relationship would have to be modified to use this criterion. Since this criterion does not conflict with the hypothesis on which the ECBW method is based, the quality of ideas would gradually improve. For example, this criterion would be useful for the sixth family in which the initial idea had a high $ISR_{i,j}$ with the OBW method although the $ISR_{i,j}$ of the offspring decreased with an increase in the generation number until the fifth generation.

Only the ECBW method produced many good ideas in the first and tenth families regardless of the initial idea due to its robustness in idea quality improvement, as shown in Figs. 7(c, d). Figure 8 shows that the ECBW method produced a couple of good ideas after the eleventh generation in these families while the $ISR_{i,j}$ with the ECBW and OBW methods were comparable until the eleventh generation. In the first and tenth families, the ECBW method produced ideas with high $ISR_{i,j}$ after successive suggestion of poor ideas at around the tenth and fifteenth generations, respectively. These reproductions correspond to the red-dashed boxes in Fig. 5(a). By contrast, the OBW method did not produce good ideas in the first family due to the successive suggestion of poor ideas after producing three reasonable ideas ($ISR_{i,j} = 2/3$) from the fifth to the eighth generations.

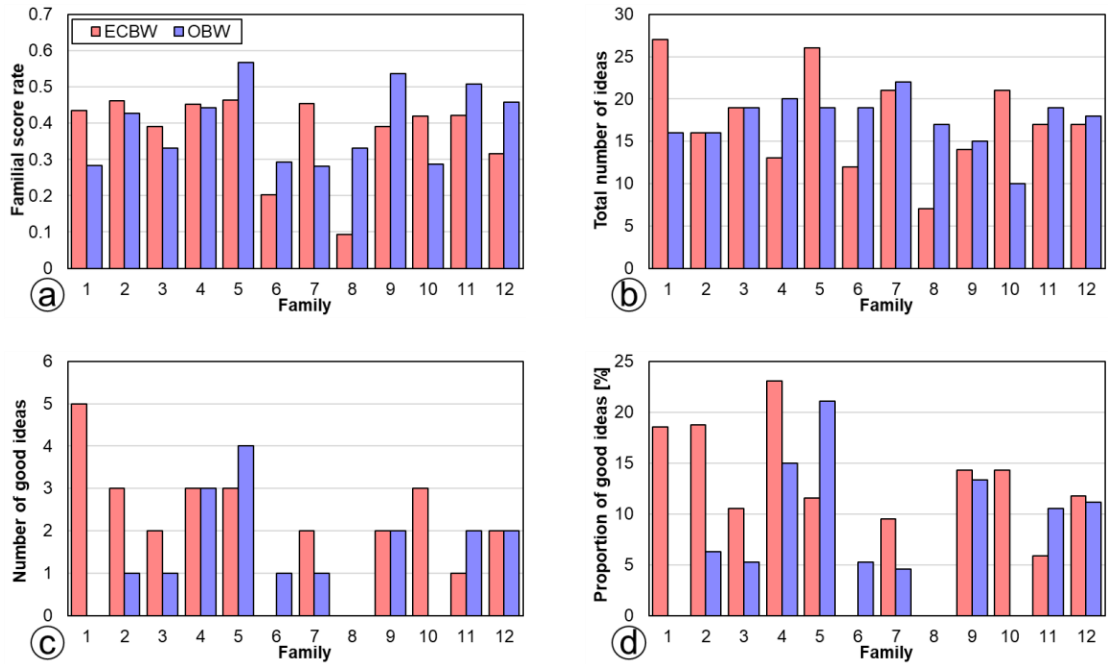

Figure 7: Performances of families at end of ideation with ECBW and OBW methods: (a) familial score rates, (b) total number of ideas, (c) number of good ideas, (d) proportion of good ideas to total ideas.

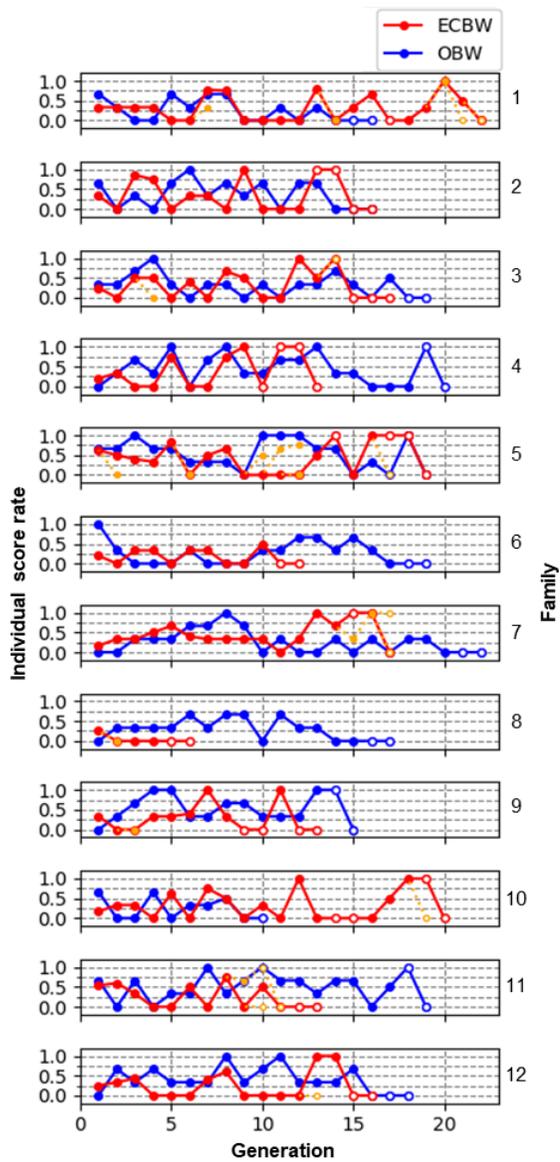

Figure 8: Individual score rate variation by generation by family.

Comparing the initial ideas in the fourth and eighth families, we see that their $ISR_{i,j}$ was low, as shown in Fig. 8, whereas the quality of the families (represented by $FSR_i$) and the proportion of good ideas (Fig. 7) were substantially different. These results mean that the $ISR_{i,j}$ of ideas and their suitability for stimulating ideas are not always correlated. If the suitability for stimulation can be quantified and used for individual selection, the proportion of good ideas should increase. Using the $ISR_{i,j}$ of an offspring as an individual selection criterion for its parent is one approach to suitability quantification. Definition of the parent-offspring relationship would have to be modified to use this criterion. Since this criterion does not conflict with the hypothesis on which the ECBW method is based, the quality of ideas would gradually improve. For example, this criterion would be useful for the sixth family in which the initial idea had a high $ISR_{i,j}$ with the OBW method although the $ISR_{i,j}$ of the offspring decreased with an increase in the generation number until the fifth generation.

Only the ECBW method produced many good ideas in the first and tenth families regardless of the initial idea due to its robustness in idea quality improvement, as shown in Figs. 7(c, d). Figure 8 shows that the ECBW method produced a couple of good ideas after the eleventh generation in these families while the $ISR_{i,j}$ with the ECBW and OBW methods were comparable until the eleventh generation. In the first and tenth families, the ECBW method produced ideas with high $ISR_{i,j}$ after successive suggestion of poor ideas at around the tenth and fifteenth generations, respectively. These reproductions correspond to the red-dashed boxes in Fig. 5(a). By contrast, the OBW method did not produce good ideas in the first family due to the successive suggestion of poor ideas after producing three reasonable ideas ($ISR_{i,j} = 2/3$) from the fifth to the eighth generations.

## 5 CONCLUSION

We have presented an evolutionary computation-assisted brainwriting for large-scale online ideation. In this method, participants not only suggest ideas but also evaluate ideas previously suggested by the other participants. The evaluation results are used in the interactive evolutionary computation to select ideas to be presented to the participants to stimulate their thinking in the brainwriting framework. The performance of the proposed method was compared with that of a simple online brainwriting method by applying them to a large-scale online ideation session with more than 30 participants. The proposed method robustly improved idea quality during the entire progress of ideation by preferentially presenting good ideas to the participants to stimulate their thinking. The online brainwriting method improved idea quality in a step-by-step manner although the successive suggestion of poor ideas disrupted the improvement. Evaluation of the ideas generated in a parent-offspring relationship supported the hypothesis on which the proposed method is based; even better ideas can be obtained if people are exposed to better stimulating ideas.